\documentstyle[twocolumn,psfig,rotating]{mn2e}
\def\lsim{~\rlap{$<$}{\lower 1.0ex\hbox{$\sim$}}}
\def\bsim{~\rlap{$>$}{\lower 1.0ex\hbox{$\sim$}}}
\def\kms{\ {\rm km\,s^{-1}}}

\def\hmpc{\ {\rm {\it h}^{-1}Mpc}}

\def\hmmpc{\ {\rm {\it h}Mpc^{-1}}}

\def\dd{{\rm d}}
\def\ln{{\rm ln}}

\def\la{\langle}
\def\ra{\rangle}
\def\vx{{\bf x}}
\def\vz{{\bf z}}
\def\vr{{\bf r}}
\def\vs{{\bf s}}
\def\vv{{\bf v}}
\def\vk{{\bf k}}

\def\vp{v_\parallel} 
\def\vkpe{{\bf k_\perp}}
\def\vkpa{{k_\parallel}}
\def\dko{{\frac{\dd^2\vkpe}{(2\pi)^2}}}
\def\dkop{{\frac{\dd^2\vkpe'}{(2\pi)^2}}}
\def\vnabla{\pmb{$\nabla$}}
\def\dj{{\rm 3D}}
\def\di{{\rm 1D}}
\def\om#1{\Omega_{#1}}

\def\pmb#1{\setbox0=\hbox{#1}%
\kern-.025em\copy0\kern-\wd0
\kern.05em\copy0\kern-\wd0
\kern-.025em\raise.0433em\box0}
\def\etal{{\it et al.\ }}
\newcommand{\op}{Ly$\alpha$\ }

\begin{document}

\title[] {Redshift distortions in one-dimensional power spectra}
\author[Desjacques {\it et al.}]{Vincent 
Desjacques and Adi Nusser \\ 
The Physics Department and
the Asher Space Research Institute, Technion, Haifa 32000, Israel \\
Email~: dvince@physics.technion.ac.il, adi@physics.technion.ac.il} 
\maketitle

\begin{abstract}
  
  We present a model for one-dimensional (1D) matter power spectra in
  redshift space as estimated from data provided 
   along individual  lines
  of sight. We derive analytic expressions for these power spectra in
  the linear and nonlinear regimes, focusing on redshift distortions
  arising from peculiar velocities.  In the linear regime, redshift
  distortions enhance the 1D power spectra only on small scales and do
  not affect the power on large scales.  This is in contrast to the
  effect of distortions on three-dimensional (3D) power spectra
  estimated from data in 3D space, where the enhancement is
  independent of scale.  For CDM cosmologies the 1D power spectra in
  redshift and real space are similar for wavenumbers $q\lsim
  0.1\hmmpc$ where both have a spectral index close to unity,
  independent of the details of the 3D power spectrum.  Nonlinear
  corrections drive the 1D power spectrum in redshift space into a
  nearly universal shape over scales $q\lsim 10\hmmpc$, and suppress
  the power on small scales as a result of the strong velocity shear
  and random motions.  The redshift space 1D power spectrum is mostly
  sensitive to the amplitude of the initial density perturbations.  Our
  results are useful in particular for power spectra computed from the
  SDSS quasars sample.

\end{abstract}

\begin{keywords}

intergalactic medium - quasars: absorption lines - cosmology: 
theory -- gravitation -- dark matter

\end{keywords}

\section {Introduction}
\label{introduction}

Measurement of distances in cosmology is only possible for nearby
cosmic objects, and even there it is plagued with observational
biases.  Velocities, however, are directly measured by means of the
Doppler redshifts. Most observational data  is therefore described in
terms of redshifts. Three dimensional galaxy surveys provide the
angular positions and the redshifts of galaxies. Absorption lines in
quasars (QSO) spectra are characterized by their redshift  coordinate
along lines of sight (LOS) to quasars.  Redshifts differ from
distances by the peculiar velocities (deviations from pure Hubble
flow) along the line of sight.  This causes systematic differences
between the spatial distribution of data in redshift and distance
space. These differences are  termed redshift distortions.  On large
scales, these redshift distortions amount to an enhancement of the
power spectrum estimated from the galaxy distributions in redshift
surveys. In a seminal paper, Kaiser (Kaiser 1987, see also Lilje \&
Efstathiou 1989, McGill \etal 1990) derived an expression which
accounts for the effect of linear peculiar motions in
three-dimensional (3D) power spectra. Since then, the impact of
redshift distortions on the 3D power spectrum (and higher order
statistics) has been studied intensively using analytic and numerical
methods to constrain $\om{\Lambda}$ and $\om{m}$ from measurements of
the distortions parameter $\beta$  (e.g. Peacock 1992; Hamilton 1993;
Peacock \& Dodds 1994; Fisher, Sharf \& Lahav 1994; Cole, Fisher \&
Weinberg 1994; Fisher \& Nusser 1996; Ballinger, Peacock \& Heavens
1996; Loveday \etal 1996; Ratcliffe \etal 1996; Taylor \& Hamilton
1996; Heavens \& Matarrese \& Verde 1998; Magira, Jing \& Suto 2000;
Kang \etal 2002)

In this paper  we will focus  on the effect of redshift distortions in
one-dimensional (1D) power spectra, as they could be computed from
data along individual lines of sight. We will also address how
nonlinear corrections affect the shape and the amplitude of the 1D
redshift space power spectrum. To model the linear and nonlinear
evolution of 1D  power spectra, we will rely on analytic formulae
which have been calibrated by means of numerical simulations
(e.g. Bardeen \etal 1986; Smith \etal 2003). The subject of this work
is timely in view of the large  sample of QSOs  now provided by the
Sloan Digital Sky Survey (SDSS).  The SDSS collaboration obtained
spectra of comoving length $\sim 80\hmpc$  for $\sim 100000$ quasars,
with spectral resolution $R\sim 2000$ in the range 3800-9200$\AA$
(e.g. Abazajian \etal 2003). These spectra  will improve significantly
the statistics of the (dimensionless) flux power spectrum $\Delta_{\rm
F}$ of the \op  forest which has been extensively used in the past
years to  constrain the linear 3D matter power spectrum
$\Delta_\dj^{\rm r}$  (e.g. Croft \etal 1998, 1999, 2002; McDonald
\etal 2000, McDonald 2003).

The paper is organized as follow. In \S\ref{basic} we present our
notation  and briefly review the  linear theory of redshift
distortions in 3D power spectra.  In \S\ref{linear1D} we apply  linear
theory   to model distortions in 1D power spectra.  In
\S\ref{nonlinear1D} we extend the analysis to include nonlinear
effects. We conclude with a discussion of our results in
\S\ref{discussion}.

\section{Basic relations}
\label{basic}

In this section we describe briefly the basic definitions and
relations relevant to our calculations of redshift distortions in 1D
power spectra.

Our notation is as follows.  The comoving real space (distance) and
redshift coordinates, both in $\kms$, are denoted by $\vx$ and $\vs$,
respectively.  The comoving peculiar velocity in $\kms$ at position 
$\vx$ is $\vv$.  The local density of a distribution of particles is 
$\rho$ and the corresponding density contrast is $\delta=\rho/\bar\rho-1$ 
where $\bar\rho$ is the mean density. The cosmological mass density and
vacuum energy parameters are $\om{m}$ and $\om{\Lambda}$ respectively.
The superscripts $r$ and $s$ will designate quantities in real and
redshift space, respectively.

\subsection{3D and 1D power spectra}

The three-dimensional Fourier transform of a 3D field $f(\vx) $ is
\begin{equation}
f^\dj_{\vk}=\int\!\!\frac{\dd^3\vx}{(2\pi)^3} 
\exp(-i \vk \cdot \vx) f(\vx) \; .
\end{equation}
If the field $f$ is given only along the $z$ axis, we then define its
1D Fourier transform at a wavenumber $q$ as
\begin{equation}
f^\di_{q}=\int\!\!\frac{\dd z}{2\pi} \exp(-i q z) f(z) \; ,
\end{equation}
where for brevity we denote a point on the z axis only by its $z$
coordinate (the $x$ and $y$ coordinates being zeros).  We can express
$f^{\rm 1D}_{q}$ in terms of $f^{\rm 3D}_{\vk}$ as (e.g. Kaiser \& 
Peacock 1991)
\begin{equation}
f^\di_{q}=\int\!\!\dko f^\dj_{\vkpa=q,\vkpe} \; ,
\label{d:2d3}
\end{equation}
where $k_\parallel$ and $\vk_\perp$ are the components of $\vk$ parallel
and perpendicular to the $z$ axis. It should be noted that this relation 
holds in real and redshift space.

Using (\ref{d:2d3}) the covariance matrices  of the 1D and 3D Fourier 
modes are given by
\begin{equation}
\la f_{q'}^{\rm s}f_q^{\rm s}\ra=\int\!\!\dko\int\!\!\dkop\,
\la f_{q',\vkpe'}^{\dj s}f_{q,\vkpe}^{\dj s}\ra\;.
\label{covariance}
\end{equation}
The 3D and 1D power spectra are given in terms of the covariance 
matrices by
$\la f_\vk^\dj f_{\vk'}^\dj\ra=(2\pi)^3
P_\dj(\vk)\delta_{\rm D}(\vk+\vk')$, and
$\la f_q f_{q'}\ra=2\pi
P_{1D}(q)\delta_{\rm D}(q+q')$. 
We work with the dimensionless power spectrum $\Delta$, which is the
density variance per logarithmic interval of wavenumber $k$.  Adopting 
the convention of Peebles (1980), we have
\begin{equation}
\Delta_\dj(k)={k^3\over 2\pi^2}P_\dj(k),~~~
\Delta_\di(k)={k\over\pi}P_\di(k)\;.
\label{p:def}
\end{equation}
Using (\ref{d:2d3}) and (\ref{covariance}) we arrive at 
the relation
\begin{equation}
\Delta_\di(q)=\frac{q}{2\pi}\int\!\!\dd^2\vkpe\,
\frac{\Delta_\dj(q,\vkpe)}{\left(q^2+\vk_\perp^2\right)^{3/2}}\; .
\label{power}
\end{equation}
This a general relation which is valid for 3D and 1D power spectra
in real and redshift space. This relation will be the basis of our 
calculation of redshift distortions in 1D power spectra. 

\subsection{Redshift distortions}

The corresponding redshift space coordinate, also in $\kms$, is
\begin{equation}
\vs=\vx +{\vp(\vx)}\;,
\label{s:def}
\end{equation}
where $\vp(\vr)=\vv\cdot\hat{\vx}$, and $\hat{\vx}=\vx/|\vx|$ is the
unit vector along the line of sight.   Assuming a one-to-one mapping
between $\vx$ and $\vs$, the density in redshift space, $\rho^{\rm s}$, 
can be related to $\rho^{\rm r}$  by means of the equation of mass 
conservation (continuity)
\begin{equation}
\rho^{\rm s}(\vs)d^3\vs=\rho^{\rm r}(\vx)d^3\vx \; .
\label{s:con}
\end{equation} Using (\ref{s:def}) and expanding
the continuity  equation  to 
first order in $\delta^{s,r} $ and $\dd \vp /\dd r$
yields 
\begin{equation}
\delta^{\rm s}(\vs)= \delta^{\rm r}(\vx)-{d\vp\over dr}(\vx)\;,
\label{s:contrast}
\end{equation}
The continuity equation (\ref{s:con}) and its linearised version
(\ref{s:contrast}) refer to the distribution of any points that have
the velocity field $\vp$. In this paper we are mainly interested in
the effect of redshift distortions in 1D matter power spectra, and the
fluctuation field $\delta^{\rm r}$  is simply the matter density
contrast $\delta^{\rm r}_m$.  The present calculation can be
extended to fluctuation fields which are closely related to the 
underlying mass  (dark matter) distribution by means of a linear 
bias factor, $b$ (e.g. Kaiser 1987). In this case, $\delta^{\rm r}=b
\delta^{\rm r}_m$ where $\delta^{\rm r}_m$  is the real space mass
density contrast. In the case of galaxy surveys,  $ b$ is the
traditional bias factor between the galaxy and mass distributions.

\subsection{Linear distortions in 3D power spectra}

As clear from equation (\ref{s:contrast}), a quantification of the
effect of redshift distortions on the properties of an observed field
$\delta^{\rm s}$ requires knowledge of the velocity field $\vv$.  
Over scales where the density contrast is small, linear theory 
yields the relation
\begin{equation}
\delta^{\rm r}=-\beta^{-1} \vnabla \cdot \vv \; ,
\end{equation}
where $\beta$ is proportional to the logarithmic derivative of the 
growth factor, $\dd\ln D_+/\dd\ln a$ (e.g. Peebles 1980). 
$\beta\approx\Omega_m(z)^{0.6}/b$ is a good approximation for a wide 
range of CDM models. Hence, at redshift $z\bsim 1$, one can safely 
assume $\beta\approx b^{-1}$.

In linear theory, the 3D power spectra  of $\delta^{\rm s}$ and
$\delta^{\rm r}$ can be related  as follows (e.g. Kaiser 1987).
Assuming potential flow, i.e. that the velocity is derived from a
potential, the linear theory relation can be expressed in Fourier
space as
\begin{equation}
\vv_\vk={i\vk\over k^2}\beta\delta^{\rm r}_\vk\;,
\label{linear1}
\end{equation}
where $\vv_\vk$ and $\delta^{\rm r}_\vk$ are the Fourier coefficients
of the velocity and real space density fields, respectively.
 
Here and throughout the paper we work in the ``distant observer''
limit, and assume that the observed region lies in the $z$ direction
(see e.g. Heavens \& Taylor 1995, Zaroubi \& Hoffman 1996 for the 
general case).  Hence, we write $\mu=\vk\cdot{\hat\vz}$, and the 
Fourier transform of eq.~(\ref{s:contrast}) leads, for a given 
wavenumber $\vk$,
\begin{equation}
\delta_\vk^{\rm s}=\delta_\vk^{\rm r}\left(1+\beta\mu^2\right)\;.
\label{linear2}
\end{equation}
The redshift space, dimensionless power spectrum $\Delta_\dj^{\rm
s}(\vk)$  is then
\begin{equation}
\Delta_\dj^{\rm s}(\vk)=\Delta_\dj^{\rm r}(\vk)
\left(1+\beta\mu^2\right)^2\;.
\label{kaiser}
\end{equation}
which is known as the Kaiser formula (Kaiser 1987).  To avoid the
proliferation of indices, we will hereafter drop indices when they 
characterise real space quantities.

\section{Linear redshift distortions in 1D power spectra}
\label{linear1D}

In this Section, we compute the 1D redshift space matter power
spectrum in the linear regime. The bias factor is therefore
$b=1$. Moreover, since the strength of linear redshift
distortions increases  with $\beta$, we choose to present  results at
redshift $z=3$. This choice is also aimed at facilitating the
comparison between theory and observations  (e.g. Croft \etal 1998,
2002; McDonald \etal 2000).

\begin{figure} 
\resizebox{0.45\textwidth}{!}{\includegraphics{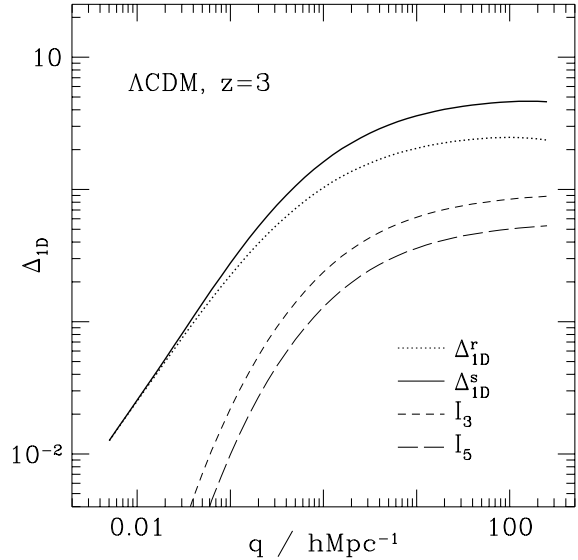}}
\caption{The real (dotted) and redshift space (solid) one-dimensional
power spectra, $\Delta_\di^{\rm r}$ and $\Delta_\di^{\rm s}$. The
function $I_3$ and  $I_5$ are also shown as short- and long-dashed
curves respectively.  The curves were computed at $z=3$ for a
$\Lambda$CDM model of spectral index $n_s=1$. On large scale, linear
redshift distortions cancel out in the  one-dimensional matter 
power spectra. }
\label{fig1}
\end{figure}

We aim at deriving an expression for the 1D power spectrum in redshift
space, $\Delta_\di^{\rm s}$, in terms of the 3D linear power spectrum
in real space, $\Delta_\dj^{\rm r}$.  To this extent we substitute
$\Delta_\dj^{\rm s}$ as given in terms of $\Delta_\di^{\rm r}$ in
eq.~(\ref{kaiser}) into the basic relation (\ref{power}) and obtain
\begin{equation}
\Delta_\di^{\rm s}(q)=q\int_q^\infty\!\!\dd k\,
\frac{\Delta_\dj^{\rm r}(k)}{k^2}
\left[1+\beta\left(\frac{q}{k}\right)^2\right]^2\;.
\label{power2}
\end{equation}
$\Delta_\dj^{\rm r}(k)$ is now the linear 3D power spectrum in real space.
Eq.~(\ref{power2}) is the counterpart of the Kaiser
relation for 1D spectra. Defining the functions $I_n(q)$ as
\begin{equation}
I_n(q)=q^n\int_q^\infty\!\!\dd k\,\frac{\Delta_\dj^{\rm r}}{k^{n+1}}\;,
\label{moment}
\end{equation}
we can write the 1D, redshift space power spectrum as follows,
\begin{equation}
\Delta_\di^{\rm s}(q)=\Delta_\di^{\rm r}(q)+2\beta I_3(q)+\beta^2 I_5(q)\;,
\label{ps1}
\end{equation}
where we used the fact that $I_1(q)$ is  the 1D power spectrum in real
space $\Delta_\di^{\rm r}$. Using the relation~(\ref{power}), it is
also possible to integrate the functions $I_n$ by part and express
$\Delta_\di^{\rm s}(q)$ as a function of $\Delta_\di^{\rm r}(q)$ alone.

\subsection{The linear PS in a $\Lambda$CDM cosmology}
\label{linexample1D}

In Fig~\ref{fig1}, we show the real and redshift space 1D power
spectra, $\Delta_\di^{\rm r}$ and $\Delta_\di^{\rm s}$, as the dotted
and solid curves respectively. They are computed using the 3D power
spectrum given by the fitting formula of Bardeen \etal (1986) for a
$\Lambda$CDM model. The cosmological parameters that we use are
listed in Table~\ref{table1}.  The $\Lambda$CDM model has a
present-day, {\it rms} of fluctuations $\sigma_8=0.9$ on scale
$8\hmpc$, and a spectral index $n_s$. A value $n_s=0.9-1$ is
consistent with the latest observations of the CMB, the large-scale
structure and the \op forest, which constrain the spectral index to be
$n_s=0.93\pm 0.03$ on scale $0.07\hmmpc$ (Spergel \etal 2003).

\begin{table}
\caption{The main parameters of the CDM models {\bf considered} in this 
paper.}
\vspace{1mm}
\begin{center}
\begin{tabular}{cccc} \hline
         & $\om{\rm m}$ & $\om{\Lambda}$ & h \\
               \hline
    $\Lambda$CDM  & 0.3    & 0.7   & 0.7 \\
    OCDM          & 0.3    & 0     & 0.7 \\      
    SCDM          & 1      & 0     & 0.5 \\
\hline\hline
\end{tabular}
\end{center}
\label{table1}
\end{table}

\subsubsection{The large scale behaviour} 

\begin{figure*} 
\resizebox{0.45\textwidth}{!}{\includegraphics{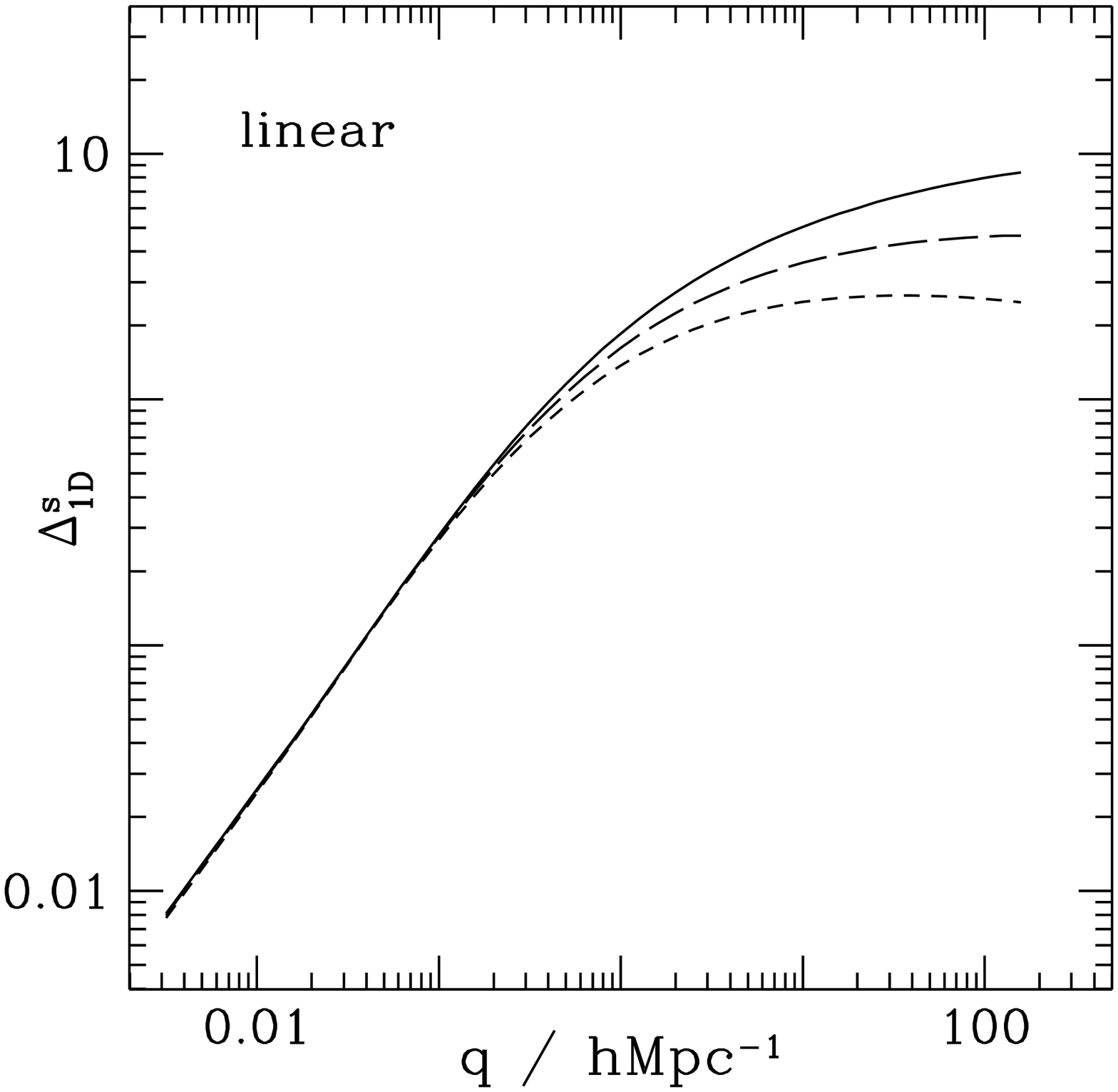}}
\resizebox{0.45\textwidth}{!}{\includegraphics{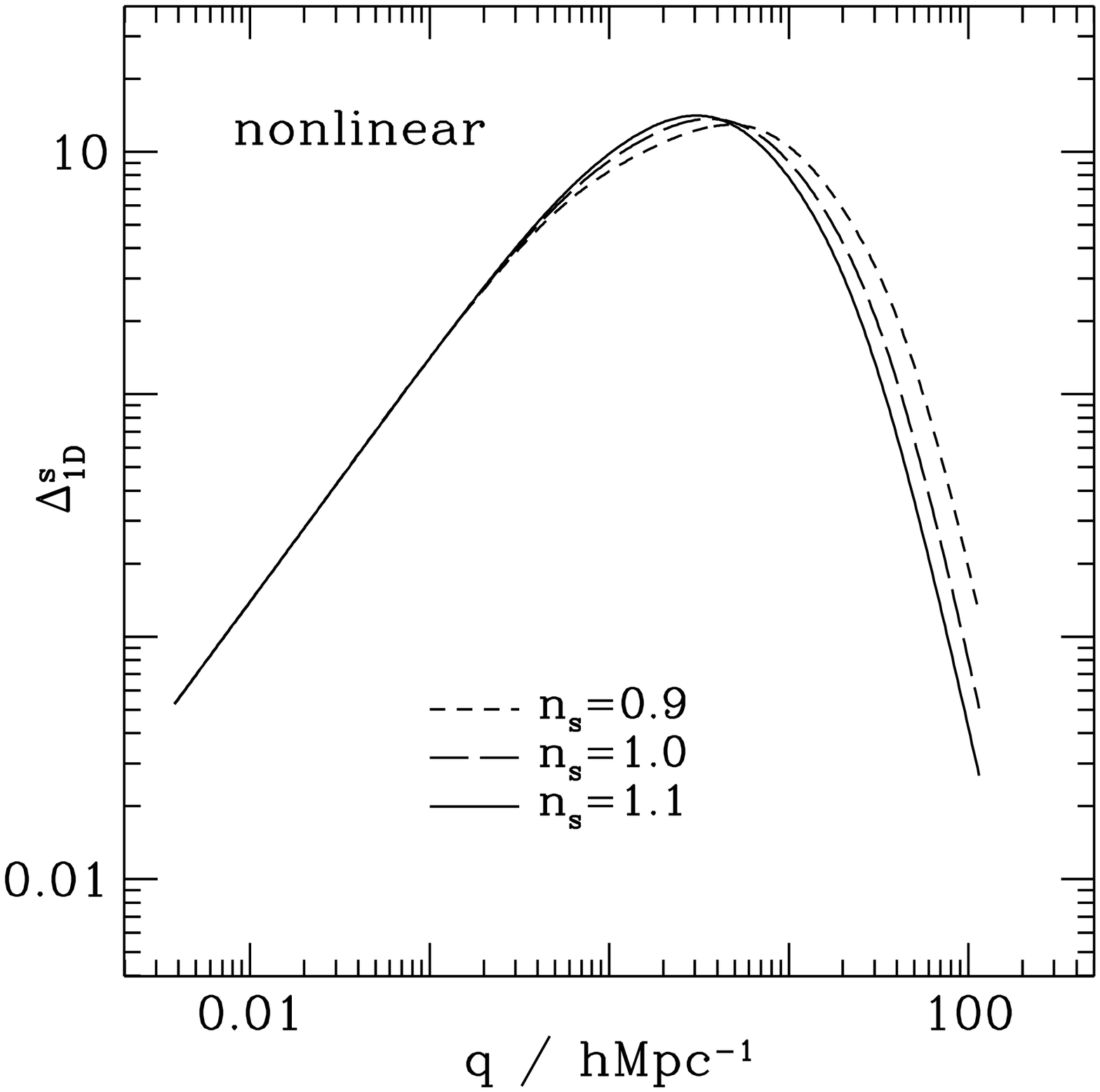}}
\caption{The linear (left panel) and nonlinear (right panel) , 1D
redshift space power spectra $\Delta_\di^{\rm s}$ for various
spectral index~: $n_s=0.9$ (short-dashed), 1 (long-dashed), and 1.1
(solid). In both panels, the curves corresponding to the power spectra
with $n_s\ne 1$ are shifted vertically  to  the  $n_s=1$ curve on
large scales. The results are   for a $\Lambda$CDM model at $z=3$.}
\label{fig2}
\end{figure*}

In Fig.~\ref{fig1} we see that redshift distortions vanish on large
scales, although one might have expected a large-scale effect from the
Kaiser formula.  Moreover, the figure shows also that the slope of the
redshift power spectrum approaches unity in the large-scale limit,
irrespective of the spectral index.  To explain this, let us examine in
more detail the large scale behaviour of each of the terms on the
r.h.s of (\ref{ps1}).  Consider first the term $\Delta_{\di}^{\rm
r}$. Using (\ref{power}), we can express it in term of the 3D power 
spectrum $\Delta_\dj^{\rm r}$ as
\begin{equation}
\frac{\Delta_\di^{\rm r}}{q}=\int_q^\infty\!\!\dd k\,
\frac{\Delta_\dj^{\rm r} (k)}{k^2} \; .
\label{moment1}
\end{equation}
In CDM cosmological models  $\Delta_\dj^{\rm r}\sim k^{n_s+3}$ and
$k^{1-n_{s}}$ on large and small scales scales respectively. Thus, the
integral in (\ref{moment1})  converges as $q$ tends to infinity so
that $\Delta_{\di}^{\rm r}\sim q$ in that  limit. Following similar
arguments, we find $I_{3}\sim q^{3}$ and $I_{5}\sim q^{n_{s}+3}$ in
that limit. Therefore, the main contribution to $\Delta_\di^{\rm s}$
on large scales is the term $\Delta_\di^{\rm r}\sim q$.

\subsubsection{The small scale behaviour}

Fig.~\ref{fig1} shows that redshift distortions enhance the power by a
factor of $\sim 2$ on scales $q\bsim 0.1\hmmpc$.  On these scales
$\Delta_\di^{\rm s}$ is very sensitive to the spectral index $n_s$. This 
is clear from the left panel of Fig.~\ref{fig2}, where the linear, 1D
redshift space power spectrum is plotted for spectral index
$n_s=0.9,1$ and 1.1.  Curves corresponding to models with $n_s\ne 1$,
were shifted vertically so that they match the model $n_s=1$ on large
scales. Since we are working in the linear regime, this matching can
also be achieved by a change in $\sigma_8$ for these models.  For
$q\bsim 0.1\hmmpc$, the amplitude of the 1D, redshift space
fluctuations increases significantly with the spectral index. Based on
this, one would hope to constrain the spectral index from measurements
on scale $q\bsim 0.1\hmmpc$. However, as we will show next the small
scale power spectrum in redshift space is strongly affected by
nonlinear effects.

\section{Nonlinear redshift distortions in 1D power spectra}
\label{nonlinear1D}

So far we considered linear perturbations for which the 3D power
spectra in redshift and real space are related by the Kaiser
relation~(\ref{kaiser}). At redshift $z\sim 3$ however, nonlinear
corrections are already important for wavenumbers $|\vk|\bsim 1\hmmpc$. 
As clearly seen from relation~(\ref{power}), nonlinearities
contaminate $\Delta_\di(q)$ at scales significantly larger than the
nonlinear scale (e.g. Zaldarriaga \& Scoccimarro \& Hui 2003).

\subsection{The nonlinear 3D power spectrum}

Given the initial power spectrum we wish to estimate the non-linear
1D power spectrum in redshift space.  Nonlinearities come in two
different ways.  First, the dynamical growth of density perturbations
is faster than linear theory prediction on small scales.  Second,
nonlinear random and coherent motions suppress the small-scale power
in redshift space.  This is because overdense regions with strong
velocity shear are seen stretched along the LOS (the ``finger of God''
effect) in redshift space. This stretching wipes out the small-scale 
power in redshift space. 

The nonlinear evolution of the 3D power spectrum in real space can be
easily described by simple fitting formulae which have been calibrated
using a large suite of N-body simulations (e.g.  Hamilton \etal 1991;
Jain \etal 1995; Peacock \& Dodds 1996; Smith \etal 2003).  We will
use hereafter the fitting formula proposed in Smith \etal (2003)
(cf. Appendix C of their paper) to model the nonlinear 3D power
spectrum given an initial linear power spectrum.

Incorporating the effect of peculiar velocities is more complicated.
The reason is the lack of a convenient relation, similar to the linear
eq.~(\ref{linear1}), between the velocity and density in the nonlinear
regime.  However, several workers in the field have investigated the
effect of peculiar motions on the 3D power spectrum in redshift space
using N-body simulations (e.g. Peacock 1992; Peacock \& Dodds 1994;
Cole \& Fisher \& Weinberg 1995; Bromley, Warren \& Zurek 1997;
Magira,  Jing \& Suto 2000; Jing \& B\"orner 2001).  It was found that
this effect can be modelled in the 3D power spectrum by multiplying
the r.h.s.  of eq.~(\ref{kaiser}) with a filtering function of the
form (Jing \& B\"orner 2001),
\begin{equation}
D[k\mu\sigma_{12}]=
\left[1+\frac{1}{2}\left(k\mu\sigma_{12}\right)^2+
\eta\left(k\mu\sigma_{12}\right)^4\right]^{-1}\;.
\label{motion2}
\end{equation}
where $\sigma_{12}$ is the LOS pairwise velocity dispersion (PVD) of
dark matter on scale $k\sim 1/r$, and $\eta$ is a constant calibrated
using numerical simulations.  For illustration, we have $\eta=0.00759$
for a $\Lambda$CDM model.  This scaling relation holds relatively well
for $k\mu\sigma_{12}\lsim 20$.  The PVD of the dark matter,
$\sigma_{12}$, can be computed directly from N-body simulation, but
the calculation is rather tedious. Mo, Jing \& B\"orner (1997) provide
a simple, physically motivated fitting formula for
$\sigma_{12}(k)$. This formula is a reasonable fit to the CDM
cosmological models examined in the present paper. Note that at large
separations, the pairwise velocity dispersion is
$\sigma_{12}(0)=\sqrt{2/3}\la v_1^2\ra^{1/2}$, where $\la
v_1^2\ra^{1/2}$ is the density-weighted {\it rms} peculiar velocity of
the dark matter. We also have typically $\la v_1^2\ra^{1/2}\approx
500-1000\kms$ in the CDM cosmological models considered here.
Therefore, we can mimic the small-scale damping caused by nonlinear
motions by combining the fitting formula~(\ref{motion2}) of Jing \&
B\"orner (2001) with the ansatz of Mo, Jing \& B\"orner (1997).
Although numerical simulations are needed to accurately calibrate
$D[k\mu\sigma_{12}]$ as a function of the cosmological model (in
particular the spectral index $n_s$) and redshift, we expect this
approximation to be valid for $n_s\simeq 1$.

\begin{figure} 
\resizebox{0.45\textwidth}{!}{\includegraphics{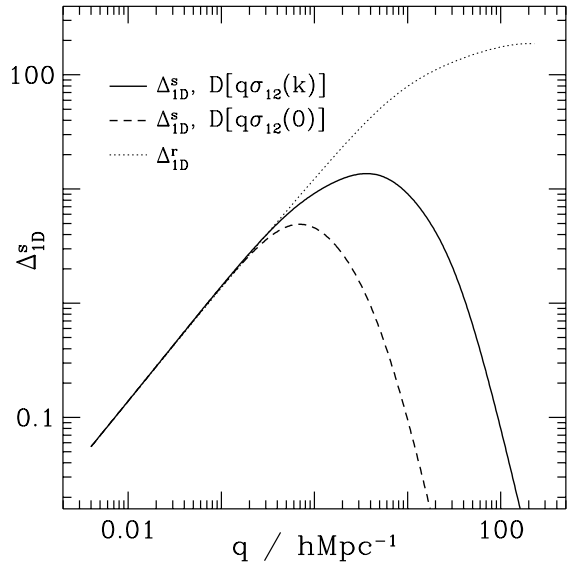}}
\caption{A comparison at $z=3$ between the nonlinear 1D, real and
redshift space power spectra, $\Delta_\di^{\rm r}$ and
$\Delta_\di^{\rm s}$, for a $\Lambda$CDM model with $n_s=1$.  The
dotted curve is $\Delta_\di^{\rm r}$. The solid curve shows
$\Delta_\di^{\rm s}$ with a damping factor $D=D[q\sigma_{12}(k)]$,
whereas the dashed curve is $\Delta_\di^{\rm s}$ with
$D=D[q\sigma_{12}(0)]$. The behaviour of $\Delta_\di^{\rm s}$ is very
sensitive to the assumed filter $D$.}
\label{fig3}
\end{figure}

\subsection{The nonlinear 1D power spectrum}

Following the previous discussion, we extent the validity of
relation~(\ref{power2}) to the nonlinear regime by computing the 1D
power spectrum $\Delta_\di^{\rm s}$ from the 3D, redshift space power
spectrum according to
\begin{equation}
\Delta_\di^{\rm s}(q)=q\int_q^\infty\!\!\dd k\,
\frac{\Delta_\dj^{\rm NL}(k)}{k^2}
\left[1+\beta\left(\frac{q}{k}\right)^2\right]^2
D\left[q \sigma_{12}(k)\right]\;.
\label{power3}
\end{equation}
In the general case, $\sigma_{12}$ is a function of $k$ and the
damping factor $D[q\sigma_{12}(k)]$ cannot be extracted from the
integral.  In Fig.~\ref{fig3} we compare at $z=3$ the nonlinear 1D,
real space power spectrum (dotted curve) with its redshift space
counterpart as computed from eq.~(\ref{power3}) (dotted curve).  The
damping factor $D$ erases the power on scale $q\bsim 1\hmmpc$. A
comparison with the 1D power spectrum $\Delta_\di^{\rm s}$ as computed
from a damping factor $D=D[q\sigma_{12}(0)]$ (long-dashed) reveals
also that $\Delta_\di^{\rm s}$ is very sensitive to the exact
behaviour of $\sigma_{12}(q)$. Finally, it should be noted that, since
the contribution from virialized dark matter haloes to the pairwise
dispersion $\sigma_{12}$ increases with time, the characteristic
wavenumber of the cutoff decreases with decreasing redshift.

\begin{figure*}
\resizebox{1.0\textwidth}{!}{\includegraphics{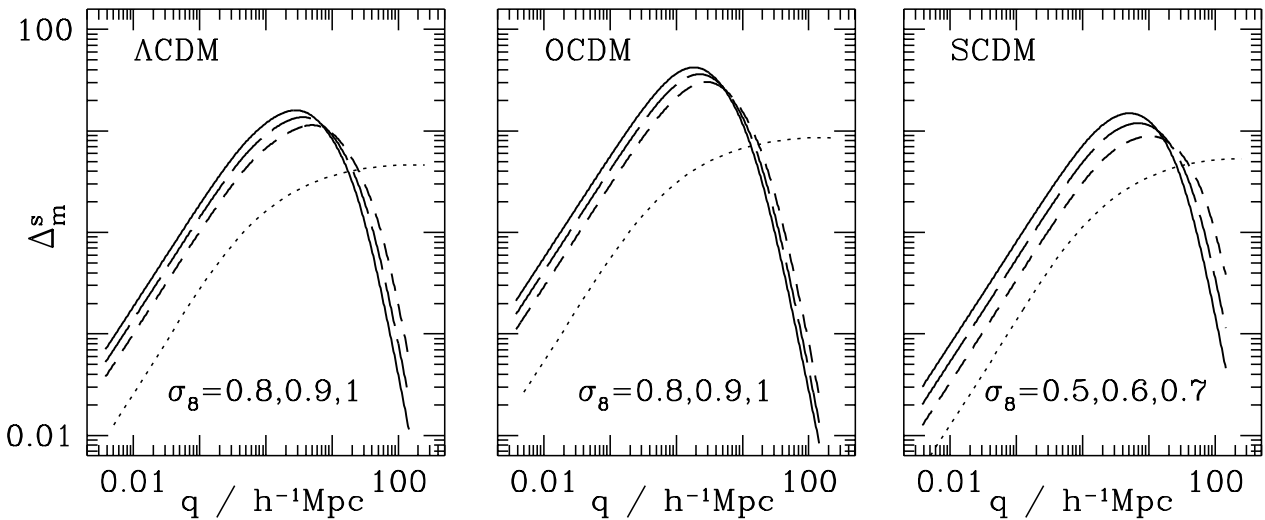}}
\caption{The nonlinear 1D, redshift space power spectrum
$\Delta_\di^{\rm s}$ at $z=3$ for the CDM models of
Table.~\ref{table1}, with spectral  index $n_s=1$. In each panel,
$\Delta_\di^{\rm s}$ is plotted for three  different $\sigma_8$ as
indicated on the Figure.  The solid (short-dashed) curve corresponds
to the highest (lowest) quoted values of $\sigma_8$.  We also show as
a dotted curve the linear $\Delta_\di^{\rm s}$ corresponding to the
nonlinear $\Delta_\di^{\rm s}$ plotted as  a long-dashed curve.  For a
given cosmological model, the amplitude of $\Delta_\di^{\rm s}$, and
the  position of its maximum are sensitive to the value of $\sigma_8$.}
\label{fig4}
\end{figure*}

In this analysis, we did not take into  account higher order terms of
the Taylor expansion of the continuity equation~(\ref{s:contrast})
(see, for example, Cole, Fisher \& Weinberg 1994 for a discussion of
this effect). However, these corrections are  implicitly included in
the low-pass band filter $D$. One could also perform a perturbative
calculation to address this issue in a more consistent  way
(e.g. Heavens, Matarrese \& Verde 1998; Viel \etal 2004). 
Perturbation theory is useful for describing distortions in 3D power 
spectra in the mildly
nonlinear regime. However, according to relation (\ref{power}), 1D 
power spectra on a given scale are contaminated by the evolution of 
3D power on all smaller scales, and one should therefore ensure that 
the 3D power spectrum decays rapidly enough on small scale before 
applying perturbation theory. In this paper we follow the practical 
approach outlined above.

\subsection{Sensitivity to the cosmological parameters}

\subsubsection{The large scale behaviour}

To illustrate the dependence of the 1D power spectra  on the
cosmological parameters, we plot on Fig.~\ref{fig4} the nonlinear 1D,
redshift space power spectrum $\Delta_\di^{\rm s}$. It is computed
from equations (\ref{power2}) and  (\ref{power3}) for various CDM
models~: $\Lambda$CDM (left panel), OCDM (middle panel) and SCDM
(right panel). The corresponding cosmological parameters are listed in
Table~\ref{table1}. For each cosmological model, we plot
$\Delta_\di^{\rm s}$ for three different values of $\sigma_8$, as
quoted in Fig.~\ref{fig4}. In each panel, the  solid (short-dashed)
shows $\Delta_\di^{\rm s}$ for the largest (lowest)  $\sigma_8$. Note
that the spectral index is set to $n_s=1$.  Fig.~\ref{fig4}
illustrates the importance of nonlinear corrections, which drive the
1D redshift space power spectra $\Delta_\di^{\rm s}$ of various CDM
models towards a power spectrum nearly insensitive to the cosmological
parameters and primordial spectral index. This is a consequence of
nonlinear gravitational dynamics, which  drive  the linear, 3D real
space power spectrum $\Delta_\dj^{\rm r}$ towards a universal power-law
spectrum of effective spectral  index $n=1.6$ (Scoccimarro  \&
Frieman 1996, see also Zaldarriaga, Scoccimarro \& Hui 2003). As a
result,   aliasing causes the effective spectral index $n_\di$ to be
insensitive to $n_s$ on scale $q\lsim 1\hmmpc$.  Fig.~\ref{fig4}
demonstrates also that aliasing substantially enhances the large-scale
normalisation of $\Delta_\di^{\rm s}$. At redshift $z=3$, the
large-scale amplitude of the nonlinear power spectrum $\Delta_\di^{\rm
s}$ is larger than its linear  counterpart by a factor 5.6, 7.9 and
4.3 for the $\Lambda$, O and SCDM models respectively.  Neglecting
nonlinear corrections would therefore lead to a severe overestimation 
of $\sigma_8$.

\subsubsection{The small scale behaviour}

As we can see from Fig.~\ref{fig4}, the nonlinear 1D, redshift space
power spectrum reaches its maximum for some $q$ in the range
$1-10\hmmpc$. On smaller scale, $\Delta_\di^{\rm s}$ features a sharp
cutoff due to nonlinear motions. The precise shape of the small-scale
cutoff depends on the assumed filter $D$. It is sensitive to the
amplitude of $\sigma_{12}(q)$ which, for a given cosmological model,
is mostly set by $\sigma_8$.  As expected, the amplitude of the power
spectrum decreases with $\sigma_8$, and $\Delta_\di^{\rm s}$ peaks at
smaller scales.  Therefore, although the nonlinear $\Delta_\di^{\rm
s}$ does not seem to change much with $\om{m}$ and $\om{\Lambda}$, its
overall shape depends on the exact value $\sigma_8$. The right panel
of Fig.~\ref{fig2} shows that the shape of the nonlinear
$\Delta_\di^{\rm s}$ is also sensitive to the spectral index
$n_s$. However, the small-scale nonlinear $\Delta_\di^{\rm s}$ is much
less sensitive to $n_s$ than that in the linear regime.  There is
obviously a degeneracy between $\sigma_8$ and $n_s$. Unfortunately,
since the fitting formulae of Mo, Jing \& B\"orner (1997)  and Jing \&
B\"orner (2001) were estimated for $n_s=1$ CDM numerical simulations,
one has to extend the validity of these approximations to $n_s\ne 1$
in order to address this issue any further.

\section{discussion}
\label{discussion}

In this paper we modelled in detail power spectra estimated from data 
given along lines of sight rather than  in 3D space.  

Redshift distortions on large scales, both in the linear and nonlinear
regimes, vanish in power spectra estimated from data given along lines
of sight.  On small scales ($q\bsim 1\hmmpc$), linear redshift
distortions enhance the power spectrum, but the nonlinear random
motions and the strong velocity shear work in the opposite direction
of suppressing the power. The net result is that the nonlinear 1D
power spectrum in redshift space falls below its counterpart in real
space on small scales. On large scales however, they both have similar
shapes, but the amplitude of the nonlinear power spectrum is
significantly higher.  The dimensionless 1D power spectrum peaks at
a scale which depends on the pairwise velocity dispersion (PVD).  The
PVD is a strong function of the normalization of the power. Therefore
the shape of $\Delta_{\di}^{\rm s}$ is set by $\sigma_{8}$, and to a
lesser extent by the spectral index $n_{s}$.  For the CDM models
considered in this paper, $\Delta_\di^{\rm s}$ peaks at $q\sim
1-10\hmmpc$ which is roughly the scale at which the flux power
spectrum $\Delta_{\rm F}$ of the \op forest peaks. Note that, on scale
$q\bsim 1\hmmpc$, the flux power spectrum is also sensitive to thermal
broadening which is another redshift space effect (e.g. Theuns, Schaye
\& Haehnelt 2000).  In a future work, we will model the \op flux
power spectrum. We will also explore the implications of the 
observations on the cosmological model.

\section{Acknowledgment}

We would like to thank Saleem Zaroubi for useful discussions, and the
anonymous referee for a careful reading of the manuscript. This 
Research was supported by the Binational Science Foundation, the 
German-Israeli Foundation for the Development of Science and Research, 
and the EC RTN network ``Physics of the Intergalactic Medium''.


\begin{thebibliography}{}
\bibitem{} Abazajian K. and the SDSS team, 2003, AJ, 126, 2081
\bibitem{} Ballinger W.E., Peacock J.A., Heavens A.F., 1996, MNRAS,
282, 877
\bibitem{} Bardeen J.M., Bond J.R., Kaiser N., Szalay A.S., 1986, ApJ,
304, 15
\bibitem{} Bromley B.C., Warren M.S., Zurek W.H., 1997, ApJ, 475, 414
\bibitem{} Cole S., Fisher K.B., Weinberg D.H., 1994, MNRAS 267, 785
\bibitem{} Cole S., Fisher K.B., Weinberg D.H., 1995, MNRAS, 275, 51
\bibitem{} Croft R.A.C., Weinberg D.H., Katz N., Hernquist L., 1998,
ApJ, 495, 44
\bibitem{} Croft R.A.C., Weinberg D.H., Pettini M., Hernquist L., Katz
N., 1999, ApJ, 520, 1
\bibitem{} Croft R.A.C., Weinberg D.H., Bolte M., Burles S., Hernquist
L., Katz N., Kirkman D., Tytler D., 2002, ApJ, 581, 20
\bibitem{} Desjacques V., Nusser A., in preparation
\bibitem{} Fisher K.B., Scharf C.A., Lahav O., 1994, MNRAS, 266, 219
\bibitem{} Fisher K.B., Nusser A., 1996, MNRAS, 279, L1
\bibitem{} Hamilton A.J.S., Kumar P., Lu E., Matthews A, 1991, ApJ,
374, L1
\bibitem{} Hamilton A.J.S., 1993, ApJ, 406, L47
\bibitem{} Heavens A.F., Taylor A.N., 1995, MNRAS, 275, 483
\bibitem{} Heavens A.F., Matarrese S., Verde L., 1998, MNRAS, 301, 797
\bibitem{} Jain B., Mo H.J., White S.D.M., 1995, MNRAS, 276, L25
\bibitem{} Jing Y.P., B\"orner G., 2001, ApJ, 547, 545
\bibitem{} Kaiser N., 1987, MNRAS, 227, 1
\bibitem{} Kaiser N., Peacock J.A., 1991, ApJ, 379, 482
\bibitem{} Kang X., Jing Y.P., Mo H.J., B\"orner G., 2002, MNRAS, 336,
892
\bibitem{} Lilje P.B., Efstathiou G., 1989, MNRAS, 236, 851
\bibitem{} Loveday J., Efstathiou G., Maddox S.J., Peterson B.A.,
1996, ApJ, 468, 1
\bibitem{} Magira H., Jing Y.P., Suto Y., 2000, ApJ, 528, 30
\bibitem{} McDonald P., Miralda-Escud\'e J., Rauch M., Sargent W.L.W.,
Barlow T.A., Cen R., Ostriker J.P., 2000, ApJ, 543, 1
\bibitem{} McDonald P., 2003, ApJ, 585, 34
\bibitem{} McGill C., 1990, MNRAS, 242, 428
\bibitem{} Mo H.J., Jing Y.P., B\"orner G., 1997, MNRAS, 286, 979
\bibitem{} Peebles P.J.E., 1980, The Large Scale Structures of the 
Universe, Princeton University Press
\bibitem{} Peacock J.A., 1992, in Martinez V., Portilla M., S\'aez D.,
eds, New insights into the Universe, Proc. Valencia summer school.
Springer, Berlin, p.1
\bibitem{} Peacock J.A., Dodds S.J., 1994, MNRAS, 267, 1020
\bibitem{} Peacock J.A., Dodds S.J., 1996, MNRAS, 280, L19
\bibitem{} Ratcliffe A., Shanks T., Broadbent A., Parker Q.A., Watson
F.G, Oates A.P., Fong R., Collins C.A., 1996, MNRAS, 281, 47
\bibitem{} Scoccimarro R., Frieman J.A., 1996, ApJ, 473, 620
\bibitem{} Smith R.E, Peacock J.A., Jenkins A., White S.D.M., Frenk
C.S., Pearce F.R., Thomas P.A., Efstathiou G., Couchman H.M.P., 2003,
MNRAS, 341, 1311
\bibitem{} Spergel D.N., Verde L., Peiris H.V., Komatsu E., Nolta
M.R., and the WMAP team, 2003, ApJS, 148, 175
\bibitem{} Taylor A.N., Hamilton A.J.S., MNRAS, 1996, 282, 767
\bibitem{} Theuns T., Schaye J., Haehnelt M.G., 2000, MNRAS, 315, 600
\bibitem{} Viel M., Matarrese S., Heavens A., Haehnelt M.G., Kim
T.-S., Springel V., Hernquist L., 2004, MNRAS, 347, L26
\bibitem{} Zaldarriaga M., Scoccimarro R., Hui L., 2003, ApJ, 590, 1
\bibitem{} Zaroubi S., Hoffman Y., 1996, ApJ, 462, 25

\end{thebibliography}
\end{document}